\documentclass[aps,prb,twocolumn,showpacs,preprintnumbers,groupedaddress,amsmath,amssymb]{revtex4-1}

\usepackage{graphicx}% Include figure files
\usepackage{dcolumn}% Align table columns on decimal point
\usepackage{bm}% bold math
\usepackage{amsmath,amssymb}
\usepackage{graphicx}
\usepackage[]{natbib}
\usepackage{color}

\usepackage{epstopdf}
\usepackage{subfigure}

\def\cm{cm$^{-1}$}
\def\sco{$\alpha$-SrCr$_2$O$_4$}
\def\cco{$\alpha$-CaCr$_2$O$_4$}

\newcommand{\weak}{$\bullet\circ\circ$}
\newcommand{\med}{$\bullet\bullet\circ$}
\newcommand{\str}{$\bullet\bullet\bullet$}

\begin{document}

\title{Raman study of magnetic excitations and magneto-elastic coupling in \sco}

\author{Michael Valentine}
\affiliation{Institute for Quantum Matter and Department of Physics and Astronomy, Johns Hopkins University, Baltimore, MD 21218, USA}

\author{Seyed Koohpayeh}
\affiliation{Institute for Quantum Matter and Department of Physics and Astronomy, Johns Hopkins University, Baltimore, MD 21218, USA}

\author{Martin Mourigal}
\altaffiliation[Present Address: ]{School of Physics, Georgia Institute of Technology, Atlanta, GA 30332, USA}
\affiliation{Institute for Quantum Matter and Department of Physics and Astronomy, Johns Hopkins University, Baltimore, MD 21218, USA}

\author{Tyrel M. McQueen}
\affiliation{Institute for Quantum Matter and Department of Physics and Astronomy, Johns Hopkins University, Baltimore, MD 21218, USA}

\author{Collin Broholm}
\affiliation{Institute for Quantum Matter and Department of Physics and Astronomy, Johns Hopkins University, Baltimore, MD 21218, USA}

\author{Natalia Drichko}\email{Corresponding author: drichko@jhu.edu}
\affiliation{Institute for Quantum Matter and Department of Physics and Astronomy, Johns Hopkins University, Baltimore, MD 21218, USA}

\author{Si\^an E. Dutton}
\altaffiliation[Present Address: ]{Cavendish Laboratory, University of Cambridge, Cambridge, CB3 0HE, UK}
\affiliation{Institute for Quantum Matter and Department of Chemistry, Princeton University, Princeton, NJ 08544, USA}

\author{Robert J. Cava}
\affiliation{Institute for Quantum Matter and Department of Chemistry, Princeton University, Princeton, NJ 08544,  USA}

\author{Turan Birol}
\affiliation{School of Applied and Engineering Physics, Cornell University, Ithaca, NY 14853, USA}
\affiliation{Department of Physics and Astronomy, Rutgers University, Piscataway, NJ 08854, USA}

\author{Hena Das}
\affiliation{School of Applied and Engineering Physics, Cornell University, Ithaca, NY 14853, USA}

\author{Craig J. Fennie}
\affiliation{School of Applied and Engineering Physics, Cornell University, Ithaca, NY 14853, USA}

\date{\today}

\begin{abstract}

Using Raman spectroscopy, we investigate the lattice phonons, magnetic excitations, and magneto-elastic coupling in the distorted triangular-lattice Heisenberg antiferromagnet \sco, which develops helical magnetic order below 43~K. Temperature dependent phonon spectra are compared to predictions from density functional theory calculations which allows us to assign the observed modes and identify weak effects arising from coupled lattice and magnetic degrees of freedom. Raman scattering associated with two-magnon excitations is observed at 20~meV and 40~meV.  These energies are in general agreement with our \textit{ab-initio} calculations of exchange interactions and earlier theoretical predictions of the two-magnon Raman response of triangular-lattice antiferromagnets. The temperature dependence of the two-magnon excitations indicates that spin correlations persist well above the N\'eel temperature.

\end{abstract}

\date{\today}
\maketitle

\section{Introduction}
\label{sec:intro}

The triangular-lattice Heisenberg antiferromagnet is a central model in frustrated magnetism. The theoretically established ground state has long-range magnetic order for any spin (including $S\!=\!1/2$), with moments ordered in a coplanar 120$^{\circ}$ structure.\cite{Huse1988,Jolicoeur1989,Capriotti1999} For sufficiently large $S$, the magnetic excitation spectrum is well captured by spin-wave theory with strong magnon interactions due to the non-collinear character of the magnetic order.\cite{Chernyshev2006,Starykh2006,Chernyshev2009,Mourigal2013}

While numerous quasi-two-dimensional triangular-lattice materials are known, their ground-state properties and magnetic excitations are often profoundly modified by exchange anisotropies and interactions beyond the nearest-neighbor Heisenberg model. Examples include the distorted triangular geometry in Cs$_2$CuCl$_4$,\cite{Coldea2003} spin-space anisotropy in Ba$_3$CoSb$_2$O$_9$,\cite{Shirata2012,Zhou2012} further neighbor exchange interactions in CuCrO$_2$ \cite{Poienar2010} and LuMnO$_3$, \cite{Lewtas2010,Oh2013} and delocalized spins in $\kappa$-(BEDT-TTF)$_2$Cu$_2$(CN)$_3$\cite{Yamashita2011} and LiZn$_2$Mo$_3$O$_8$.\cite{Sheckelton2012,Mourigal14}

\cco\ and \sco\ belong to yet another family of triangular-lattice Heisenberg antiferromagnets. Their orthorhombic crystal structures comprise two inequivalent Cr$^{3+}$ sites that form distorted-triangular layers of $S\!=\!3/2$ ions in the $bc$ plane of the $Pmmn$ space-group.\cite{Pausch1974Sr,Pausch1974Ca} Their low-temperature properties attracted attention due to the development of an incommensurate helical magnetic order below $T_{\textrm{N}}\!\approx\!43$~K,\cite{Chapon2011,Toth2011,Dutton2011} multiferroic behavior,\cite{Singh2011,Zhao2012} and unconventional spin dynamics.\cite{Toth2012,Wulferding2012,Schmidt2013} In \sco, non-monotonic changes in lattice parameters were observed around $T\!\approx\!100$~K by synchrotron X-ray powder diffraction,\cite{Dutton2011} and a small electric polarization ($P\!\leq\!0.4\ \mu$Cm$^{-2}$) was detected below $T_{\textrm{N}}$ under a poling electric field.\cite{Zhao2012} Symmetry analysis for the currently accepted nuclear and spin structures indicates linear magneto-electric effects are forbidden while quadratic terms are allowed.\cite{Chapon2011,Singh2011} The lattice and spin dynamics of \sco\ is thus of particular interest to search for possible magneto-vibrational effects and lattice distortions beyond the reported paramagnetic $Pmmn$ space group.

In this work, we present comprehensive Raman scattering results from \sco\ single crystals and address the interplay between lattice dynamics and magnetism. Raman scattering is a valuable tool in studies of frustrated magnetism due to its sensitivity to local structure and symmetry and to magnetic exchange interactions through optical phonons and two-magnon scattering,\cite{Fleury1968} respectively. In some cases, Raman scattering has proven more sensitive to lattice distortions than synchrotron X-ray diffraction measurements,\cite{Kant09a,Nielson2013} and thus it is well suited to reveal the effects of weak magneto-elastic coupling.

To interpret our results, we performed density functional theory (DFT) calculations that provide theoretical values for the phonon frequencies and their corresponding eigenvectors and for the magnetic exchange interactions. This comparison allows assignment of spectral features to specific phonons and identification of small lattice distortions that precede magnetic ordering by following the temperature dependence of the Raman spectra. Furthermore, using the magnetic exchange interactions obtained \textit{ab-initio}, we compared the observed magnetic excitation spectrum with theoretical predictions for the magnetic Raman response of distorted triangular-lattice antiferromagnets.\cite{PerkinsTriangular,Vernay2007,Perkins2013}

This paper is organized as follows. Section \ref{sec:exp} contains technical details associated with our single-crystal synthesis, Raman scattering measurements, and DFT calculations. Section \ref{sec:phonons} discusses the lattice dynamics of \sco\ and contains a comparison of the observed phonon Raman spectra with DFT results. Section \ref{sec:magnons} presents our magnetic Raman scattering results along with \textit{ab-initio} calculations of the nearest-neighbor magnetic exchange interactions in \sco.

\section{Experimental and theoretical methods}
\label{sec:exp}

\subsection{Crystal growth and sample preparation}

Stoichiometric amounts of SrCO$_3$ (Alfa Aesar, 99.99\% purity) and Cr$_2$O$_3$ (Alfa Aesar, 99.97\% purity) were thoroughly ground together and then synthesized under an atmosphere of 95\% Ar + 5\% H$_2$ at 1300~$^\circ$C for 10~h. The powder was then sealed into a rubber tube, evacuated using a vacuum pump, and compacted into a rod 6~mm in diameter and 70~mm in length using a hydraulic press under an isostatic pressure of 70~MPa. After removal from the rubber tube, the rods were sintered at 1500~$^\circ$C for 12~h in a 1~bar static argon atmosphere. \sco\ single crystals of approximately 4~mm in diameter and 50~mm in length were grown from the feed rods in a four mirror optical floating zone furnace (Crystal Systems Inc. FZ-T-12000-X-VPO-PC) with four 3~kW xenon lamps. Growths were carried out under 1~bar of static high purity argon at a growth rate of 10~mm/h with rotation rates of 20~rpm for the growing crystal and 3~rpm for the feed rod. Powder X-ray diffraction data taken from the crushed single crystals was consistent with the structure of pure \sco.

We used X-ray Laue back reflection to orient several of the grown single crystals for Raman scattering measurements. The first set of oriented crystals were cut using a diamond saw and cleaved to obtain flat $bc$ plane surfaces (triangular-lattice) of high optical quality. Note that the orthorhombic structure of \sco\ stems from the Sr$^{2+}$ positions and the resulting small displacement of the first Cr$^{3+}$ site away from the higher symmetry position which corresponds to an undistorted triangular-lattice. As a consequence, the $bc$ cut contains three distinct orthorhombic domains for which the $\boldsymbol{b}$ and $\boldsymbol{c}$ axes are rotated by $\pm60^\circ$ with respect to that of a reference domain which means it was not possible to distinguish $\boldsymbol{b}$ from $\boldsymbol{c}$ during crystal alignment. A second set of oriented crystals were cut to obtain flat $ab$ plane and $ac$ plane surfaces. Due to the macroscopically indistinguishable $\boldsymbol{b}$ and $\boldsymbol{c}$ axes, these cuts correspond to a mixture of $ab$ and $ac$ orientations designated by $ab+ac$ in the following.

\subsection{Raman measurements}

Raman scattering spectra were measured in micro-Raman and macro-Raman configurations using a Jobin-Yvon T64000 triple monochromator Raman spectrometer. Micro-Raman spectra of \sco\ were measured using an Olympus microscope coupled to the spectrometer with a laser probe diameter of approximately 2~$\mu$m for the spectral range from 100~\cm\ (12~meV) to 650~\cm\ (81~meV) with a resolution of 2~\cm\ (0.25~meV). For macro-Raman scattering measurements in the 50~\cm\ (2.5~meV) to 650~\cm\ (81~meV) spectral range, we used collecting optics coupled with the macro-chamber of the same spectrometer with the diameter of the probe about 50~$\mu$m. The 514.5~nm line of a Spectra-Physics Ar$^+$-Kr$^+$ laser was used for excitation light.

For low temperature Raman measurements, the crystals were mounted on the sample holder of a Janis ST-300 $^4$He flow cryostat using silver paint.  The temperature of the sample was estimated by comparing intensities of Stokes and anti-Stokes Raman spectra at 300 and 250~K. For micro-Raman measurements 1.0~mW of laser power was used which led to heating of the sample by approximately 20~K. This power was reduced to reach lower temperatures. Macro-Raman measurements used 10~mW of laser power resulting in approximately 10~K of heating.  All Raman spectra were corrected for Bose-Einstein temperature effects.

Measurements were done in a backscattering geometry with $\boldsymbol{e}_i$ (electric field vector of the incident light) and $\boldsymbol{e}_s$ (electric field vector of the scattered light) laying in the $bc$ plane ($xx$ and $xy$ polarizations) for temperatures ranging from 300 to 14~K and with $\boldsymbol{e}_i$ and $\boldsymbol{e}_s$ in the $ab+ac$ plane ($zz$ and $xz+yz$ polarizations) at room temperature. While the $bc$ crystal surface was very high quality, the $ab+ac$ surface was not ideal which reduced the intensity of the corresponding spectra and lead to leakage between the $zz$ and $xz+yz$ polarizations. With the size of structural domains within the $bc$ plane of approximately 25-50~$\mu$m in each direction, the micro-Raman measurements with a probe of 2~$\mu$m diameter allowed measurements to be performed for a single domain in the $bc$ plane.

Additionally, the spectra were measured using both 514.5 and 488.0~nm excitation lines of a Coherent Ar$^+$ laser with a Jobin-Yvon U1000 double monochromator Raman spectrometer equipped with a photomultiplier tube detector with resolution from 3~\cm\ (0.4~meV) to 10~\cm\ (1.2~meV).  The sample was maintained at temperatures between 300 and 16~K using a custom built Janis cold finger $^4$He cryostat with the sample fixed on the cold finger by silver paint.

\subsection{Density functional theory calculations}

We used Kohn-Sham DFT as implemented in the Vienna Ab-initio Simulation Package (VASP) code.\cite{Kresse1996,Blochl94} A $4\times8\times8$ k-point grid and a 500~eV energy cutoff were used to reach good convergence of structural and response properties.  The local-density approximation was employed to estimate the exchange-correlation part of the energy functional. Phonon frequencies were calculated using the frozen phonons methods with symmetry adapted modes. We considered only collinear magnetic structures due to computational expense and ignored spin-orbit coupling. DFT+$U$ as developed by Liechtenstein \textit{et al.}\cite{Liechtenstein95} was used to take into account the strong correlations associated with Cr$^{3+}$ $d$-orbitals. The values of the on-site Hubbard $U$ and the intra-atomic Hund's coupling used were 3.0~eV and 0.9~eV, respectively. We found that orthorhombic differences in nearest-neighbor magnetic exchange $J$ is insensitive to $U$ in the range from 2~eV to 6~eV.  To determine symmetry properties, the Isotropy Software Package\cite{Stokes2007} and the Bilbao Crystallographic Server\cite{AroyoKirov2006,AroyoPerezMato2006,Aroyo2011,Tasci2012} were used. Visualization for Electronic and Structural Analysis (VESTA) software was used for visualization and calculation of bond lengths and angles.\cite{Momma2008}

\section{Phonon spectrum}
\label{sec:phonons}

\subsection{Raman active phonons}
\label{sec:RamActPhonons}

\begin{table}[t]
\begin{tabular}{|c|c|c|}
\hline
Element & Wyckoff Pos. & \, Raman Representation  \,  \\ \hline
Sr1 &   2$b$  &   $A_g$ + $B_{2g}$ + $B_{3g}$   \\ \hline
Sr2 &   2$a$  &   $A_g$ + $B_{2g}$ + $B_{3g}$   \\ \hline
Cr1 &   4$f$  &   \, 2$A_g$ + 1$B_{1g}$ + 2$B_{2g}$ + 1$B_{3g}$ \,   \\ \hline
Cr2 &   4$c$  &  Inactive \\  \hline
O1  &   4$f$  &   2$A_g$ + 1$B_{1g}$ + 2$B_{2g}$ + 1$B_{3g}$   \\ \hline
O2  &   4$f$  &   2$A_g$ + 1$B_{1g}$ + 2$B_{2g}$ + 1$B_{3g}$   \\ \hline
O3  &   8$g$  &   3$A_g$ + 3$B_{1g}$ + 3$B_{2g}$ + 3$B_{3g}$   \\ \hline
\end{tabular}
\caption{Wyckoff positions and Raman active vibrations for \sco.}
\label{symm}
\end{table}

As briefly outlined in Sec.~\ref{sec:intro}, the orthorhombic structure of \sco\ comprises edge-sharing CrO$_6$ octahedra organized in the $bc$ plane of the $Pmmn$ space-group. Magnetic Cr$^{3+}$ ($S\!=\!3/2$) ions form distorted triangular layers (see Fig.~\ref{CrLayer}) stacked along $\boldsymbol{a}$ and separated by parallel lines of Sr$^{2+}$ cations. A Rietveld refinement of the $T\!=\!100$ K neutron powder diffraction pattern yields two distinct Cr$^{3+}$ sites per unit cell with fractional coordinates $\boldsymbol{r}_1$ (Cr1) and $\boldsymbol{r}_2$  (Cr2) and Wyckoff positions $4c$ and $4f$, respectively.\cite{Dutton2011}  The lattice symmetry appears preserved for temperatures below $T_{\textrm{N}}\!\approx\!43$~K with $\boldsymbol{r}_1=(0.5049,0.25,0.4975)$ and $\boldsymbol{r}_2=( 0.5,0.5,0)$ at $T\!=\!12$~K. The low-temperature structure thus displays four distinct nearest-neighbor Cr--Cr distances varying by less than $\leq\!0.5\%$ around the average distance $\bar{d}\!=\!2.94$~\AA.

Symmetry analysis for the space-group and atomic positions of \sco\ yields 36 Raman active modes, given in Table~\ref{symm}, with Raman tensors
\begin{eqnarray}
A_g &=&
\begin{pmatrix}
a & 0 & 0 \\
0 & b & 0 \\
0 & 0 & c
\end{pmatrix},\label{eqn:ramtens} \\
B_{1g} = \nonumber
\begin{pmatrix}
0 & 0 & 0 \\
0 & 0 & d \\
0 & d & 0
\end{pmatrix},
B_{2g} &=&
\begin{pmatrix}
0 & 0 & e \\
0 & 0 & 0 \\
e & 0 & 0
\end{pmatrix},
B_{3g} =
\begin{pmatrix}
0 & f & 0 \\
f & 0 & 0 \\
0 & 0 & 0
\end{pmatrix}.
\end{eqnarray}

\begin{figure}
    \includegraphics[width=8cm]{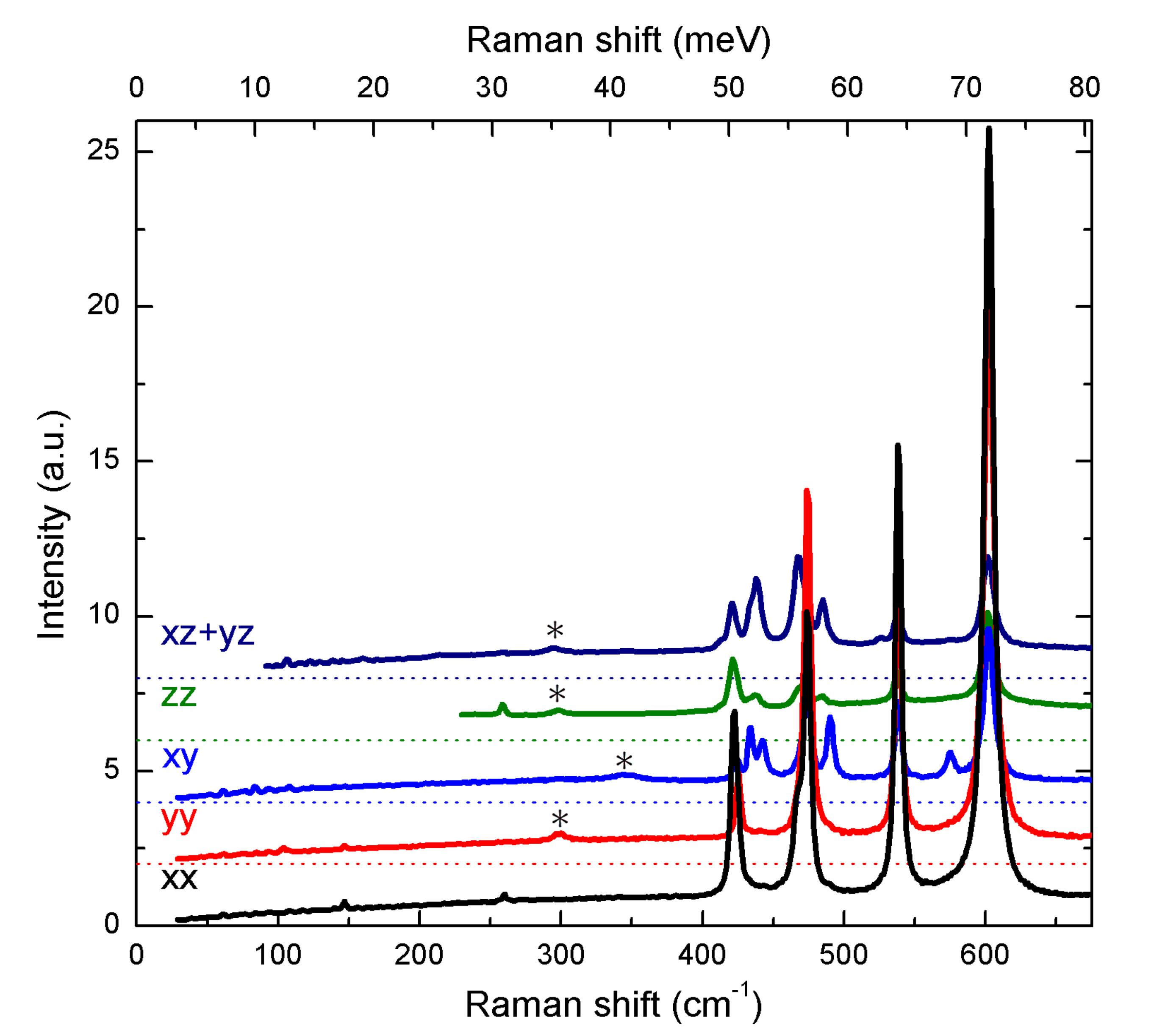}\\
    \caption{Room temperature micro-Raman spectra of \sco\ in $xx$, $yy$, $xy$, $zz$ and $xz+yz$ polarizations. Phonons showing strong spin-phonon coupling are indicated by an asterisk ($\ast$).} \label{scoRoom}
\end{figure}

The room-temperature polarized Raman spectra of \sco\ are presented in Fig.~\ref{scoRoom}. As is typical for transition-metal oxides, modes below 300~\cm\ are primarily associated with vibrations of the metal atoms. Phonons in the range from 400 to 650~\cm\ involve oxygen vibrations of the CrO$_6$ octahedra. In the measurements done on $ab+ac$ surface we could not separate $yz$ and $xz$ polarizations ($B_{1g}$ and $B_{2g}$ modes).

In Table~\ref{Phonons}, we compare the experimentally observed phonon frequencies with those obtained by DFT calculations. Out of the 36 calculated Raman active modes, 28 are experimentally observed.  The majority of the missing modes have $B_{1g}$ and $B_{2g}$ symmetry which is explained by the low signal from the  $ab+ac$ surface of the crystal. Other discrepancies are likely due to overlap of weak peaks with stronger modes. In particular, many of the oxygen vibrations are close in frequency so there is ambiguity in their assignment. Overall, the calculated phonon frequencies agree remarkably well with the experimentally obtained values. This demonstrates that DFT calculations provide a good description of the lattice dynamics of \sco\ and by inference its derivatives such as \cco.

\begin{table}
{\small
\begin{tabular}{|c|c|c||c|c|c|}
\hline
\multicolumn{3}{|c||}{Experiment} & \multicolumn{3}{c|}{Theory} \\ \hline
\cm  & Pol. & Int. & \cm & Sym. & Atomic Motions \\ \hline\hline
81   &   $xy$      & \weak &   83  &   $B_{3g}$  &  Sr1($y$) + Sr2($y$)           \\ \hline
102  &   $yy$      & \weak &  100  &   $A_g$     &  Sr1($x$) + Sr2($x$)           \\ \hline
144  & $xx$, $yy$  & \weak &  145  &   $A_g$     &  Sr1($x$) + Sr2($x$)           \\ \hline
144  &   $xy$      & \weak &  145  &   $B_{3g}$  &  Sr1($y$) + Sr2($y$)          \\ \hline
163  & $xz + yz$   & \weak &  146  &   $B_{2g}$  &  Sr1($z$) + Sr2($z$) + Cr1($z$)  \\ \hline
     &             &       &  221  &   $B_{2g}$  &  Sr1($z$) + Sr2($z$)           \\ \hline
259  & $xx$, $zz$  & \med  &  249  &   $A_g$     &  Cr1($z$) + O2($z$) + O3($y$)    \\ \hline
     &             &       &  263  &   $B_{2g}$  &  Cr1($z$) + O2($xz$) + O3($y$)   \\ \hline
295  & $xz + yz$   & \weak &  $\ast$308  &   $B_{2g}$  &  Cr1($x$) + O1($x$) + O2($x$)    \\ \hline
298  &  $yy$, $zz$ & \weak &  $\ast$310  &   $A_g$     &  Cr1($x$) + O1($x$) + O2($x$)    \\ \hline
     &             &       &  $\ast$372  &   $B_{1g}$  &  Cr1($y$) + O3($z$)            \\ \hline
346  &   $xy$      & \weak &  $\ast$373  &   $B_{3g}$  &  Cr1($y$) + O3($xz$)           \\ \hline
413  & $xz + yz$   & \weak &  409  &   $B_{2g}$  &  O1($z$) + O3($xz$)             \\ \hline
421  & $xz + yz$   & \weak &  424  &   $B_{1g}$  &  O3($yz$)                    \\ \hline
422  & $xx$, $zz$  & \str  &  413  &   $A_g$     &  O1(z) + O3($xyz$)           \\ \hline
     &             &       &  426  &   $B_{2g}$  &  O2($xz$) + O3($y$)            \\ \hline
433  &   $xy$      & \med  &  426  &   $B_{3g}$  &  O3($yz$)                    \\ \hline
425  &   $yy$      & \str  &  428  &   $A_g$     &  O2($z$) + O3($y$)             \\ \hline
434  & $xz + yz$   & \weak &  440  &   $B_{1g}$  &  O1($y$) + O2($y$)             \\ \hline
438  & $xz + yz$   & \med  &  453  &   $B_{2g}$  &  O1($x$) + O3($xy$)            \\ \hline
442  &   $xy$      & \med  &  442  &   $B_{3g}$  &  O1($y$) + O2($y$)             \\ \hline
\scalebox{0.8}{$\blacksquare$}467  &   $xx$    & \med  &  454  &   $A_g$     &  O1($x$) + O3($xy$)            \\ \hline
     &             &       &  462  &   $B_{1g}$  &  O2($y$) + O3($x$)             \\ \hline
     &             &       &  462  &   $B_{3g}$  &  O3($x$)                     \\ \hline
468  & $xz + yz$   & \str  &  463  &   $B_{2g}$  &  O1($z$) + O2($xz$) + O3($x$)    \\ \hline
\scalebox{1.0}{$\blacktriangledown$}474  &  $xx$, $yy$ & \str  &  469  &   $A_g$     &  O1($xz$) + O2($xz$) + O3($x$)   \\ \hline
485  & $xz + yz$   & \med  &  473  &   $B_{1g}$  &  O3($xyz$)                   \\ \hline
490  &   $xy$      & \med  &  476  &   $B_{3g}$  &  O3($yz$)                    \\ \hline
527  & $xz + yz$   & \weak &  515  &   $B_{2g}$  &  O1($x$) + O2($x$) + O3($yz$)    \\ \hline
\scalebox{1.0}{$\blacktriangle$}538  & $xx$, $yy$  & \str  &  528  &   $A_g$     &  O1($z$) + O2($x$) + O3($yz$)    \\ \hline
     &             &       &  569  &   $B_{1g}$  &  O1($y$) + O2($y$)             \\ \hline
575  &   $xy$      & \weak &  571  &   $B_{3g}$  &  O1($y$) + O2($y$)             \\ \hline
     &             &       &  580  &   $A_g$     &  O1($xz$) + O2($z$) + O3($z$)    \\ \hline
     &             &       &  583  &   $B_{2g}$  &  O1($xz$) + O2($x$) + O3($xz$)   \\ \hline
     &             &       &  592  &   $B_{2g}$  &  O1($z$) + O2($xz$) + O3($yz$)   \\ \hline
602  & $xx$, $yy$  & \str  &  600  &   $A_g$     &  O1($x$) + O2($xz$) + O3($x$)    \\ \hline
\end{tabular}
}
\caption{Comparison of the experimentally observed frequency, polarization, and intensity of phonon peaks with corresponding DFT calculations of Raman-active phonons. The experimental intensity ranges from weak (\weak) to strong (\str). The calculations are performed in presence of a ferromagnetic spin ordering. The major contributions to the atomic motions are listed in the last column of the table, and a more detailed account can be found in the Appendix along with the dependence of the calculated phonon frequencies on $U$ and the magnetic configurations. Calculated modes designated by $\ast$ show significant spin-phonon coupling. The temperature dependence of the modes marked with \protect\scalebox{0.8}{$\blacksquare$}, $\blacktriangledown$, and $\blacktriangle$ is presented in Fig.~\ref{fit}.}
\label{Phonons}
\end{table}

To elucidate potential magneto-elastic coupling, phonon calculations were performed imposing different collinear ferromagnetic and antiferromagnetic spin structures, and the resulting variations in phonon frequencies between these different states was assessed. Our calculations reveal that out of the 36 Raman active modes, four display a frequency shift associated with spin-phonon coupling of 15~\cm\ or more, they are marked by $\ast$ in Tab.~\ref{Phonons}. In our experiments, we observe three of these modes: 295~\cm\ $B_{2g}$ (calc. 308~\cm), 298~\cm\ $A_g$ (calc. 310~\cm),  and 346~\cm\ $B_{3g}$. Indeed, these modes are the only modes that show significant changes in frequency and width approaching $T_\textrm{N}$.  Bands associated with these phonons show full-width at half-maximum (FWHM) of $>$10~\cm\ at 300~K, significantly wider than the thermal-broadening limited FWHM of 4~\cm\ found for non-coupled phonons.

\subsection{Temperature-dependent changes in phonons}
\label{SectionTPhonons}

\begin{figure}
    \includegraphics[width=8cm]{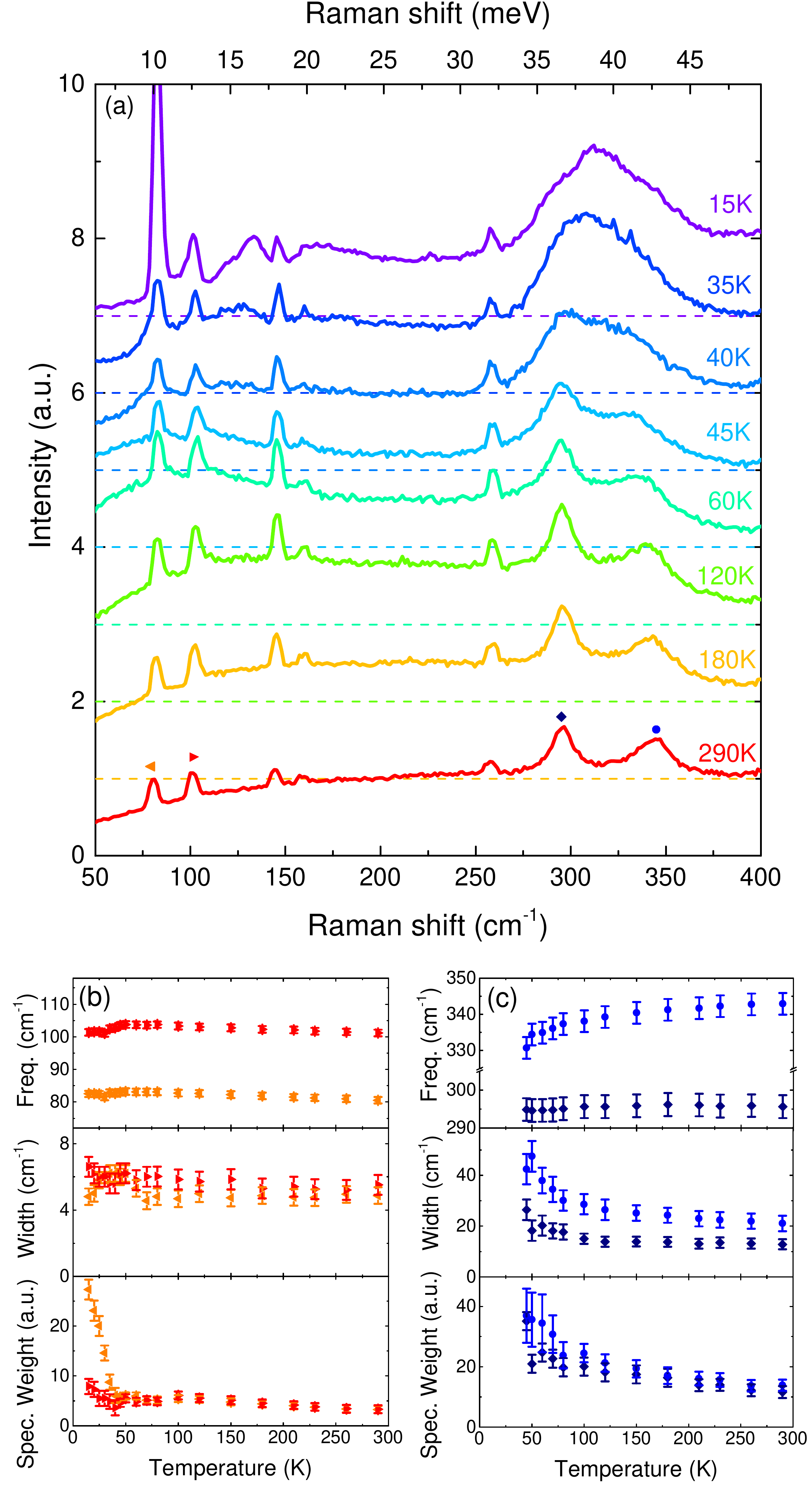}\\
    \caption{(a) Unpolarized Raman spectra of \sco\ at selected temperatures between 290 and 15~K below 400~\cm. Spectra are shifted along $y$ axis for clarity. $y=0$ is shown for each spectrum by a dashed line. On cooling down from 290~K, the magnetic background starts to increase and forms two wide features which become narrower and shift to higher frequencies below $T_{\textrm{N}}$. (b) Temperature dependence of the frequency, width, and intensity for the 81~\cm\ $B_{3g}$ and 102~\cm\ $A_{g}$ phonons associated with Sr movement; (c) Temperature dependence of phonon frequency, width, and intensity for the 298~\cm\ $A_g$ and 346~\cm\ $B_{3g}$ phonons which show considerable coupling to the spin system according to the DFT calculations.}
    \label{LT}
\end{figure}

An absence of major changes in phonon spectrum of \sco\ on cooling down to 15~K is in agreement with the observation of $Pmmn$ space group in the whole studied temperature range by X-ray and neutron powder diffraction.\cite{Dutton2011} In Fig.~\ref{LT}, we report the temperature dependence of the \sco\ unpolarized Raman spectrum upon cooling from 290 to 15~K. The largest changes with temperature, especially around $T_\textrm{N}$, are expected from the phonons which show magneto-elastic coupling. According to the fits shown in Fig.~\ref{LT}(c), the 298~\cm\ $A_g$ and 346~\cm\ $B_{3g}$ phonons show broadening on approaching $T_\textrm{N}$, and the 346~\cm\ mode shows a decrease in frequency. Both of the phonons can only be distinguished in the spectra at temperatures above $T_\textrm{N}$ while below it they mix with the two-magnon feature centered at 320~\cm, discussed in further detail in Section~\ref{MRaman}.

\begin{figure}[t!]
    \includegraphics[width=8cm]{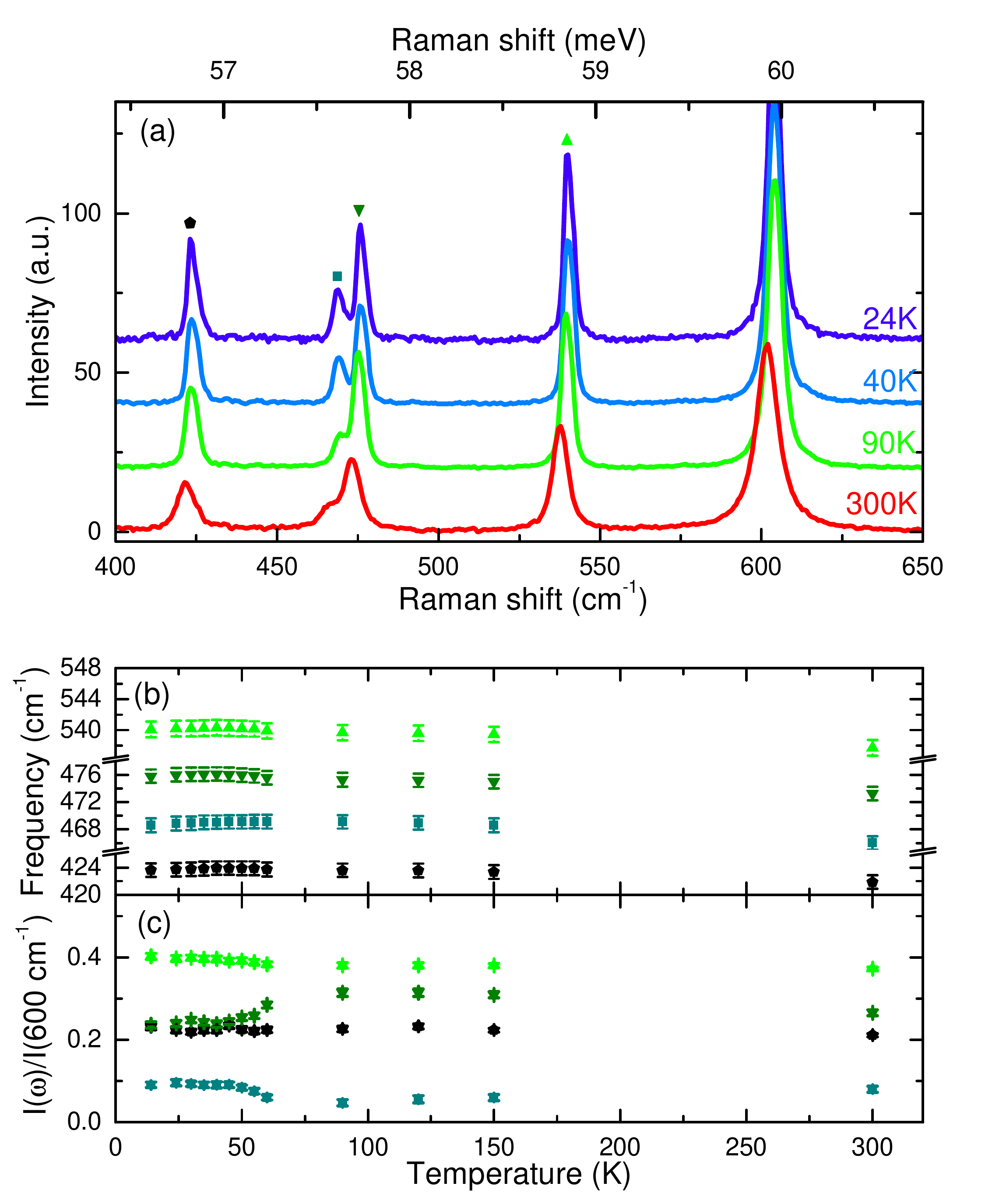}\\
    \caption{(a) Temperature dependence of the Raman spectra of \sco\ in $xx$ polarization at frequencies between 400 and  650~\cm, where oxygen-related phonons are observed. The spectra are shifted along y axis for clarity. (b) Frequencies of selected oxygen-related phonons, see the marking symbols in (a) panel. (c) Integrated intensities of these phonons normalized by the intensity of the 602~\cm\ phonon.  A redistribution of intensities occurs in the temperature range between 90 and $T_{\textrm{N}}\!\approx\!43$~K.}
    \label{fit}
\end{figure}

Along with the changes in the behavior of the phonons coupled to the spin system, we observe changes for some of the phonons which do not show substantial magneto-elastic coupling in the calculations. A dramatic increase in intensity of the 82~\cm\ $B_{3g}$ Sr phonon is observed for temperatures below $T_N$ (Fig. \ref{LT} (b)). In addition, some changes are observed for oxygen phonons (see Fig.~\ref{fit}). A spectral weight redistribution occurs  between the 467 and 474~\cm\ $A_g$ modes, and a weak change is also observed for the 538~\cm\ $A_g$ phonon.  In Fig.~\ref{fit}(b), we show the temperature dependence of their integral intensities normalized by the 602~\cm\ phonon. For temperatures below approximately 90~K, the 467~\cm\ peak doubles in intensity relative to the 474~\cm\ peak with the trend continuing until roughly $T_{\textrm{N}}\!\approx\!43$~K, below which the relative intensities remain constant. This behavior contrasts with an absence of changes observed for the 422~\cm\ phonon, which is plotted for reference in Fig.~\ref{fit}(b).

The changes in the phonons which are not coupled to the magnetic system could be due to weak variation in the structural parameters with temperature.  Indeed, the inter-plane lattice spacing $a$ decreases upon cooling from 300~K before increasing weakly for temperatures below 100~K, and an inflection point is observed in the temperature dependence of all three lattice parameters at the magnetic ordering temperature $T_{\textrm{N}}\!\approx\!43$~K\cite{Dutton2011}. Additionally, the weak electrical polarization observed below $T_\textrm{N}$\cite{Zhao2012} can influence the phonon intensities.

\section{Magnetic excitations}
\label{sec:magnons}

\subsection{Calculations of magnetic exchange interactions}
\label{meicalcs}

Originating from almost structurally perfect triangular layers of $S\!=\!3/2$ ions, the collective magnetism of \sco\ is particularly interesting. A Curie-Weiss fit to the magnetic susceptibility for temperatures above $T$=150~K indicates overall strong antiferromagnetic interactions between Cr$^{3+}$ spins with a Weiss constant $\Theta_{\textrm{W}}\!\approx\!-596$~K.\cite{Dutton2011} Below $T_{\textrm{N}}\!\approx\!43$~K, the onset of long-range magnetic order is indicated by sharp $\lambda$-anomaly in the specific-heat \cite{Dutton2011,Zhao2012,Hardy2013} and the concomitant appearance of an incommensurate magnetic Bragg peak in neutron diffraction indexed by the propagation vector ${\boldsymbol k}\!=\!(0,0.322,0)$.~\cite{Dutton2011} The magnetic structure is an incommensurate \textit{spin helix} with spins in the $ac$ plane. This is distinct from the $120^\circ$ ground-state of the nearest-neighbor triangular-lattice Heisenberg antiferromagnet.

\begin{table*}
\centering
\begin{tabular}{|c|c|c||c|c||c|c|c||c|c|c||}
\hline
\multicolumn{3}{|c||}{} & \multicolumn{5}{c||}{\sco} & \multicolumn{3}{c||}{\cco} \\ \cline{4-8} \cline{9-11}
\multicolumn{3}{|c||}{} & \multicolumn{2}{c||}{Experiment \cite{Dutton2011} $T\!=\!12$~K }
                        & \multicolumn{3}{c||}{Theory (This work)}
						& \multicolumn{3}{c||}{Experiment \cite{Toth2011,Toth2012} $T\!\leq\!5$~K} \\ \hline
& Direct exchange & Superexchange & Distance & Bond & Distance & Bond             & $J$   & Distance & Bond             & $J$      \\
& Chromiums       & Oxygens        & (\AA)    & Angle ($^\circ$) & (\AA)    & Angle ($^\circ$) & (meV) & (\AA)    & Angle ($^\circ$) & (meV)    \\ \hline
$J_1$ & Cr2--Cr2 & O1 & 2.936 & 94.7 & 2.936 & 95.5 & 4.22 & 2.907 & 94.5 & 8.6  \\ \cline{3-3} \cline{5-5}  \cline{7-7} \cline{10-10}
      &          & O2 &       & 95.1 &       & 96.3 &      &       & 94.7 &      \\ \hline
$J_2$ & Cr1--Cr1 & O3 & 2.938 & 93.5 & 2.940 & 94.2 & 7.15 & 2.911 & 92.7 & 9.1  \\ \cline{3-3} \cline{5-5}  \cline{7-7} \cline{10-10}
      &          & O3 &       & 93.5 &       & 94.2 &      &       & 92.7 &      \\ \hline
$J_3$ & Cr1--Cr2 & O3 & 2.932 & 92.4 & 2.927 & 92.7 & 5.70 & 2.889 & 91.1 & 11.8 \\ \cline{3-3} \cline{5-5}  \cline{7-7} \cline{10-10}
      &          & O1 &       & 95.4 &       & 96.0 &      &       & 94.1 &      \\ \hline
$J_4$ & Cr1--Cr2 & O3 & 2.954 & 94.7 & 2.959 & 95.3 & 3.02 & 2.939 & 94.1 & 5.8  \\ \cline{3-3} \cline{5-5}  \cline{7-7} \cline{10-10}
      &          & O2 &       & 96.6 &       & 97.0 &      &       & 95.9 &      \\ \hline
\end{tabular}
\caption{Comparison of the experimental and calculated Cr--Cr distances, Cr--O--Cr bond angles, and resulting nearest-neighbor magnetic exchange interactions $J_i$ for \sco and \cco. Theoretical magnetic exchange interactions are from \textit{ab-initio} calculations. Definitions of the $J_i$'s are given in Fig. \ref{CrLayer}.  Bond angles for \cco\ were determined from Ref. \onlinecite{Toth2011} using neutron powder diffraction measurements for atomic positions and synchrotron X-ray diffraction measurements for lattice parameters.}
\label{exchange}
\end{table*}

To understand the microscopic origin of these magnetic properties, we examined the magnetic exchange interactions of \sco\ using DFT calculations. In our treatment, exchange constants are obtained by calculating the total energy for different collinear spin configurations and fitting the results to an Ising model. The lattice was allowed to relax while preserving the overall $Pmmn$ symmetry. The results, presented in Tab.~\ref{exchange}, are compared with experimental results for \cco.

\begin{figure}[b!]
  \includegraphics[width=6cm]{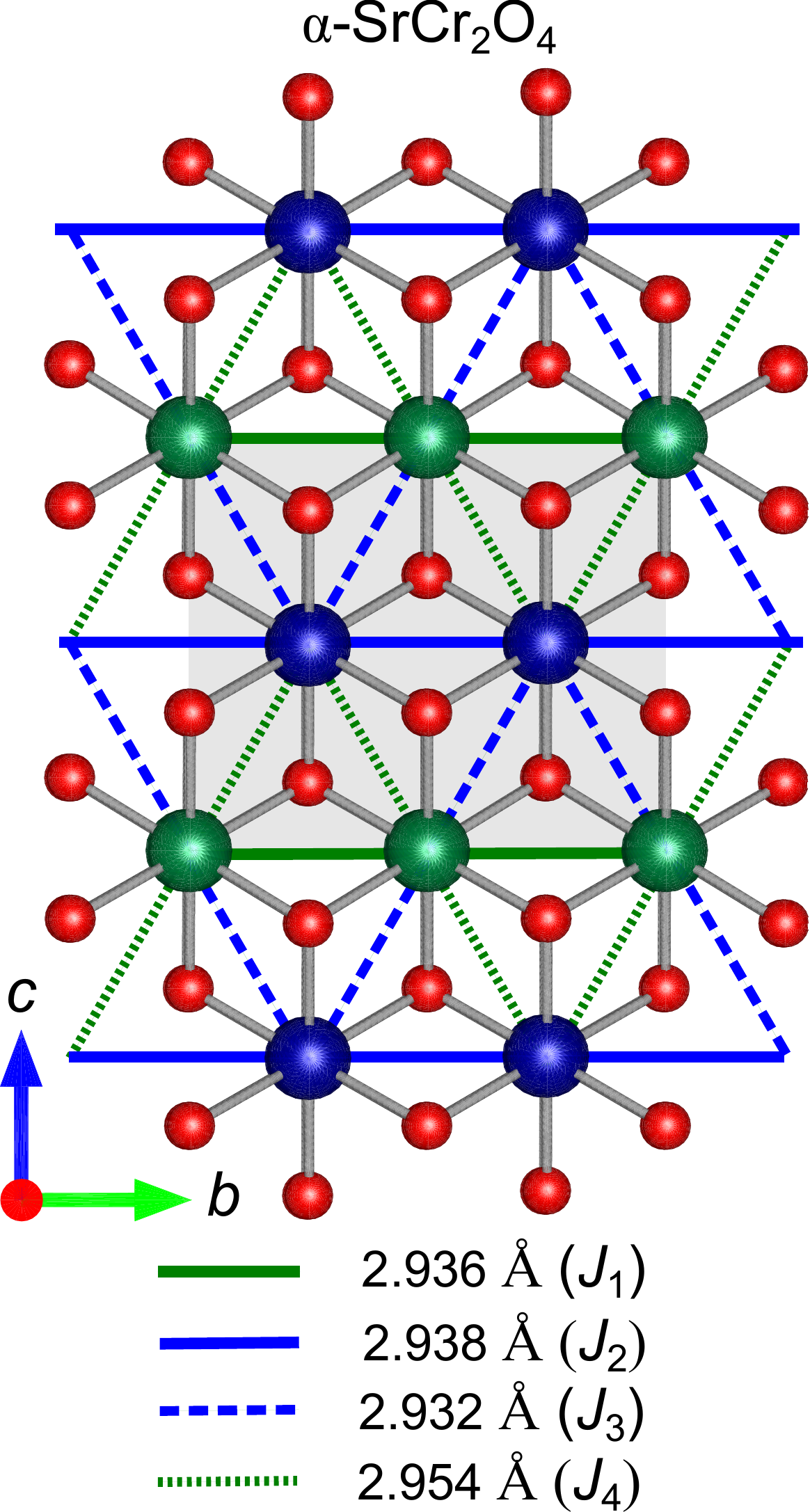}\\
  \caption{The $bc$ plane of \sco\ which consists of Cr$^{3+}$ and O$^{2-}$ ions. Blue and green spheres denote the two different Cr$^{3+}$ sites and red spheres denote O$^{2-}$ anions. There are four inequivalent Cr--Cr nearest neighbor distances, leading to the differing magnetic exchange constants $J_i$ marked in the figure. The values shown for the Cr--Cr distances were determined by neutron powder diffraction measurements at 12~K.\cite{Dutton2011}}\label{CrLayer}
\end{figure}

Although the Cr$^{3+}$ ions form an almost perfect triangular-lattice in terms of their spacings, the two inequivalent Cr$^{3+}$ sites lead to four different magnetic exchange interactions, labeled $J_1$ to $J_4$ in Fig.~\ref{CrLayer}. The DFT results indicate very strong variations in the exchange constants, with $\Delta J_i/\bar{J}\geq25\%$, even though corresponding Cr--Cr distances, vary by less than 0.5\%. Both direct cation-cation exchange interactions and superexchange through oxygen contribute to these exchange interactions. It appears that the different Cr--Cr distances cannot alone explain the large differences in exchange constants, and this suggests contributions from Cr--O--Cr superexchange play a role. For example, the value of $J_2$ is the largest where the Cr--Cr distance is second largest. The corresponding angles in the Cr--O--Cr superexchange paths are close to 90$^\circ$ where superexchange interactions are very sensitive to bond angles. Interestingly, the same tendency of the large differences between $J$'s for a nearly triangular-lattice is observed in \cco, see the experimental values in the Table III.

Overall, the theoretical estimate for the strength of magnetic interactions in \sco\  corresponds to a Weiss temperature of $\Theta_{\textrm{W}}\!=\!-583$~K ($J_{\textrm{av}} = 5.0$~meV) which compares remarkably well with the experimental value $\Theta_{\textrm{W}}\!=\!-596$~K. While this agreement might be coincidental given the well known effects of neglecting higher order terms in the high temperature expansion, we show below that the calculated $J_{\textrm{av}}$ is also consistent with the two-magnon spectrum observed experimentally by Raman scattering.

\subsection{Magnetic Raman scattering}
\label{MRaman}

\begin{figure}
  \includegraphics[width=8cm]{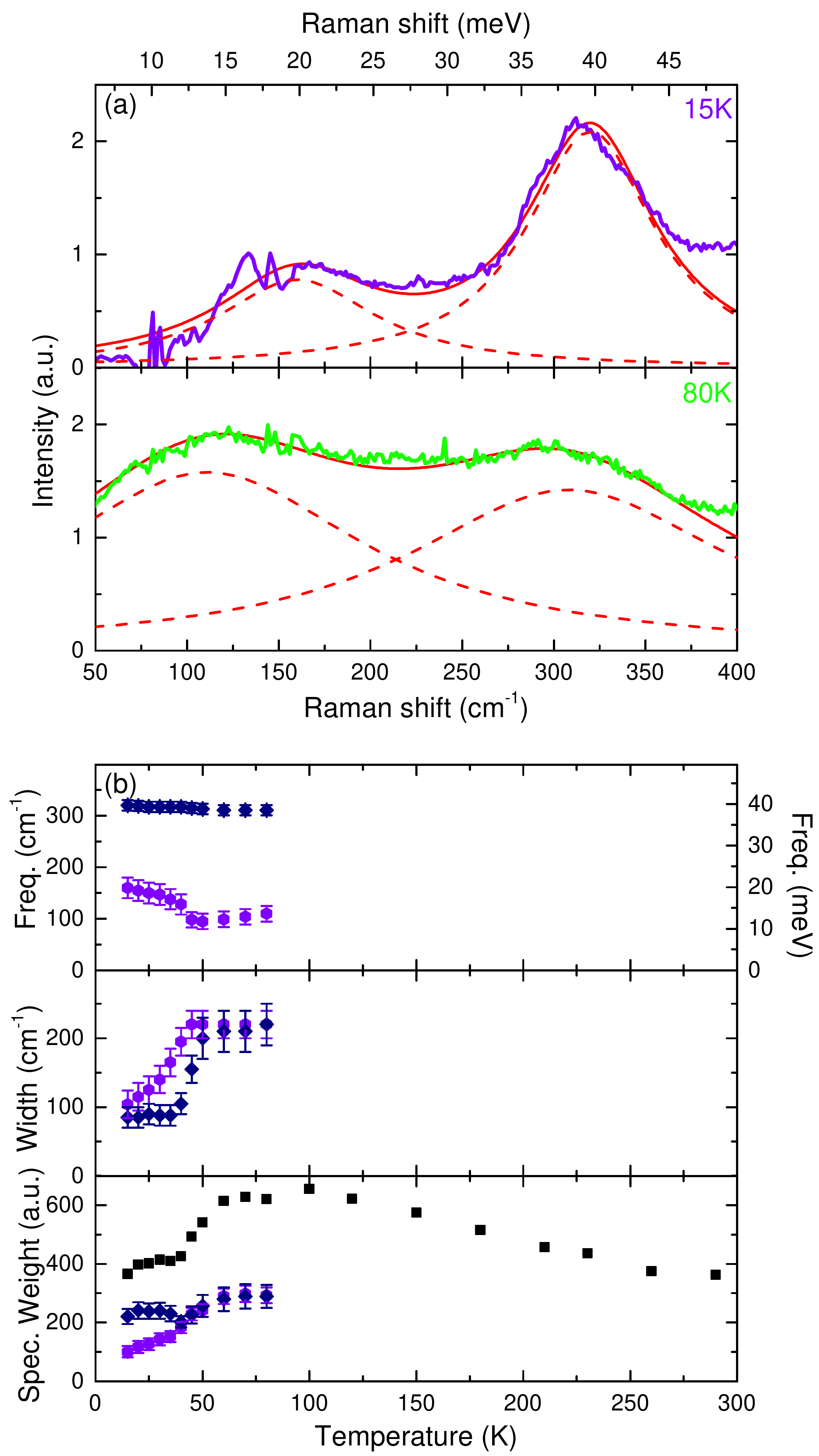}\\
  \caption{(a) Unpolarized Raman spectra of \sco\ below $T_{\textrm{N}}$ (15~K, upper panel) and above $T_{\textrm{N}}$ (80K, lower panel) with phonons extracted. Fit of the two-magnon features by Lorentzians (dashed lines) and the resulting fitting curve are shown. (b) Temperature dependence of the positions (upper panel), widths (middle panel), and spectral weights (lower panel) of the maxima of two-magnon excitations received from the fit of the spectra. The parameters are shown by diamonds (40~meW feature) and circles (20~meV feature). In the lower panel (Spectral weight), black squares present the integrated intensity $\int_{50}^{400}I(\omega)d\omega$ of the total magnetic background over the frequency range from 50 and 400~\cm.}   \label{magnonfits}
\end{figure}

In addition to phonons, below 400 \cm\ in the low-temperature spectra of \sco\ measured in the $bc$ plane (see Fig.~\ref{LT}) we observe two broad features.  At temperatures below $T_{\textrm{N}}$, these two broad peaks are centered at $\approx 20$~meV (160~\cm) and $\approx 40$~meV (320~\cm). The 40~meV peak is relatively narrow and asymmetrically skewed towards higher frequencies, while the 20~meV peak is broader and weaker. On increasing the temperature above $T_{\textrm{N}}$, both features broaden and shift to lower energies of approximately 12~meV (100~\cm) and 38~mW (310~\cm), respectively. These features are shown in greater detail in Fig.~\ref{magnonfits} in the spectra with the major phonon features extracted for temperatures below ($T$=15~K) and above ($T$=80~K) $T_{\textrm{N}}\!\approx\!43$~K. The broad and asymmetric lineshape, energy, and temperature dependence of these two bands suggest they originate from two-magnon Raman scattering.

Our experimental results can be compared with theoretical predictions for Raman excitations of triangular-lattice antiferromagnets based on the exchange scattering process\cite{Fleury1968} which is generally described by the operator $\hat{R}=\boldsymbol{S}_i J_{ij} \boldsymbol{S}_j (\boldsymbol{e}_{i}\cdot\boldsymbol{\delta}_{ij})(\boldsymbol{e}_{s}\cdot\boldsymbol{\delta}_{ij})$ where $\boldsymbol{\delta}_{ij}$ is the vector connecting neighboring sites $i$ and $j$ from different magnetic sublattices, $\boldsymbol{e}_{i}$ and $\boldsymbol{e}_{s}$ are the electric field vectors of the incident and scattered radiation, $\boldsymbol{S}_{i}$ and $\boldsymbol{S}_{j}$ are spin operators for the two different sites, and $J_{ij}$ is the magnetic exchange interaction. A crude estimation of the position of a two-magnon excitation in a collinear antiferromagnet is given by $J(2Sz-1)$, where $S$ is the spin value and $z$ is the coordination number,\cite{Cottam} which yields 17$J$ for a $S\!=\!3/2$ triangular-lattice.  Using the calculated $J_{\textrm{av}}\!=\!5.0$~meV value for \sco, the position of the observed magnetic excitations corresponds to $\approx4.0J$ and $\approx8.0J$, much lower than the prediction for a collinear antiferromagnet.

The above discrepancy can be attributed to the non-collinear character of the magnetic order in \sco. Calculations of the Raman response for a triangular-lattice antiferromagnet \cite{PerkinsTriangular,Vernay2007} predict a softening of the two-magnon excitations compared to the collinear square-lattice case. These models discuss a simple case of one magnetic lattice site per unit cell while \sco\ shows two different Cr$^{3+}$ atoms per unit cell. Nevertheless, we consider it suitable to compare our results to these simpler models since the magnetic order observed in \sco\ is non-collinear and incommensurate with the lattice. Two prominent features are expected in the Raman spectra originating from singularities in the two-magnon density of states. Ref.~\onlinecite{Vernay2007} calculates how magnetic excitations move to lower frequencies on increasing frustration through a transformation from square to triangular lattice. In the case of a slightly distorted triangular lattice and after rescaling to $S=3/2$, this results in features centered around 4.5$J$ and 6.0$J$. In Ref.~\onlinecite{PerkinsTriangular}, the positions of the magnetic excitations in the isotropic triangular lattice case are 4.5$J$ and 7.2$J$, although smoothed by magnon-magnon interactions.

Our experimental results show that in agreement to the theoretical predictions two peaks associated with two-magnon excitations are observed for a triangular antiferromagnet. The experimentally observed energies of the features, approximately 20~meV and 40~meV, or 4.0$J_{\textrm{av}}$ and 8.0$J_{\textrm{av}}$ are close to but do not exactly coincide with the theoretically predicted values. This disagreement can be explained by a simplicity of the theoretical model compared to the structure of \sco. To the best of our knowledge, magnetic Raman scattering spectra for the particular distortion of the triangular-lattice relevant for \sco\ has not been calculated, and there may be additional effects associated with the anisotropy.  In particular, the calculations of Raman spectra for an isotropic triangular lattice in Ref.~\onlinecite{PerkinsTriangular} predict that the lower energy feature should have a greater spectral weight. Our results in the ordered state below $T_{\textrm{N}}$ contrast with that prediction as the high-energy 40~meV excitation is more intense and shows a  distinct line shape compared to the weaker and broader 20~meV feature. Contrary to the square-lattice case where magnetic excitations are predicted to only occur in the $B_{1g}$ and  $B_{2g}$ polarizations,\cite{DevereauxHackl} excitations for a perfect triangular-lattice are predicted to be equally intense in the $A_{1g}$  and $B_{1g}$ channels.\cite{PerkinsTriangular,Vernay2007} Unfortunately, the presence of different orthorhombic domains in \sco\ prevents us from analyzing the polarization of the magnetic Raman spectra.

We follow the temperature dependence of the magnetic Raman scattering in \sco. To the best of our knowledge, at this point there is no published theoretical description of two-magnon scattering for triangular antiferromagnet at finite temperatures. While the real shape of the Raman spectra is defined by the two-magnon density of states,\cite{PerkinsTriangular,Cottam} we fit both observed maxima by Lorentzian band shapes to estimate their positions, widths, and intensities (see Fig.~\ref{magnonfits} (b)). Even though the positions of the two-magnon excitations in \sco\ are much lower than in non-frustrated antiferromagnets, the temperature dependence of the features across $T_{\textrm{N}}$ is similar to the observations for non-frustrated 3D and 2D materials.\cite{Cottam,Yasuda2005,Wintel1995,Ignatenko2009,DevereauxHackl} For both features associated with two-magnon excitations, the line width increases with increasing temperature above $T_\textrm{N}$, and the features shift to lower frequencies. The spectral weight of the magnetic excitations increases above $T_\textrm{N}$, also following the tendency observed for collinear non-frustrated 3D antiferromagnets.\cite{Choi2008,Cottam} We can follow the two bands as separate features up to $T\!\approx\!100$~K (Fig.~\ref{magnonfits}(b)).  Above $T\!\approx\!100$~K, the spectral weight of magnetic excitations starts to decrease, and the two basic features widen to form a magnetic background which decreases further on temperature increase, but is present up to room temperature. To illustrate that we plot the spectral weight of the whole magnetic background $\int_{50}^{400}I(\omega)d\omega$ as a function of temperature in Fig.~\ref{magnonfits} (b), lower panel (black squares), together with the  spectral weight of the two-magnon features received from the fit. The presence of magnetic excitations in 3D collinear antiferromagnets was observed in Raman scattering up to about 4$T_{\textrm{N}}$.\cite{Cottam} For \sco, the persistent 2D magnetic correlations above $T_{\textrm{N}}$ are expected given the estimated $\Theta_{\textrm{W}}\!=\!-596$~K which differs significantly from $T_{\textrm{N}}\!\approx\!43$~K due to frustration effects.

It is interesting to compare our magnetic Raman scattering results for \sco\ to those for \cco.\cite{Wulferding2012} In the latter compound, broad peaks at 5.5, 20, and 32~meV were observed and interpreted as two-magnon excitations. For both compounds the magnetic background increases on cooling the samples from room temperature, however the \cco\ compound does not show narrower bands and a decrease of the spectral weight of the magnetic background below $T_{\textrm{N}}$. The shape of the spectra below $T_{\textrm{N}}$ is different from our results on \sco, while being similar at temperatures above $T_{\textrm{N}}$. At 10~K in $RL$ polarization the data of Ref.~\onlinecite{Wulferding2012} shows a broad higher-frequency 33~meV feature with a lower intensity than the low-frequency 5.5~meV peak, in agreement with calculations for the isotropic triangular lattice.\cite{PerkinsTriangular} As a whole, the energy and line shape of the low-frequency magnetic excitations in \sco\ differ more from the theoretical predictions of Refs.~\onlinecite{PerkinsTriangular,Vernay2007} than those of \cco. This could be an indication of a more pronounced variation of nearest-neighbor magnetic interactions in \sco, as suggested by Table \ref{exchange}.

\section{Conclusions}

We presented an experimental Raman study of lattice and magnetic excitations in the anisotropic triangular antiferromagnet \sco\ along with DFT calculations for the phonon spectra and magnetic exchange constants. The calculations show an excellent agreement with the observed phonon spectra and allow us to assign all of the observed modes. The magnetic exchange interactions obtained \textit{ab-initio} show large variations as a function of subtle changes in the nearest-neighbor Cr--Cr distances. This suggests that the Cr--O--Cr superexchange mechanism plays a significant role in the nearest-neighbor exchange interactions. The average theoretical value $J_{\textrm{av}}\!=\!5.0$~meV is in excellent agreement with the Weiss constant extracted from high-temperature susceptibility data.

We detected two peaks in the magnetic Raman spectrum at approximately 20 and 40~meV which are resultant from two-magnon excitations. An observation of two peaks is close to that  predicted for two-magnon Raman scattering from triangular-lattice Heisenberg antiferromagnets, while their position at approximately 4.0$J_{\textrm{av}}$ and 8.0$J_{\textrm{av}}$ is near the expected theoretical energies. We observe a narrowing and high-frequency shift of the excitation below $T_{\textrm{N}}\!\approx\!43$~K, and both features are distinguishable up to $\approx80$~K.  Coupling between magnetic and structural degrees of freedom in \sco\ is indicated by a change in the phonons close to the frequencies of the two-magnon feature at 40~meV.

\begin{acknowledgements}

We gratefully acknowledge fruitful discussions with O. Tchernyshyov. Work at IQM was supported by the US Department of Energy, office of Basic Energy Sciences, Division of Material Sciences and Engineering under grant DE-FG02-08ER46544. Work at Cornell was supported by the NSF Grant No. DMR-1056441.

\end{acknowledgements}

\appendix*
\section{}
\subsection{Eigendisplacements of the Raman active modes from first principles}

\begin{figure*}
 \begin{center}
 \includegraphics[width=16cm]{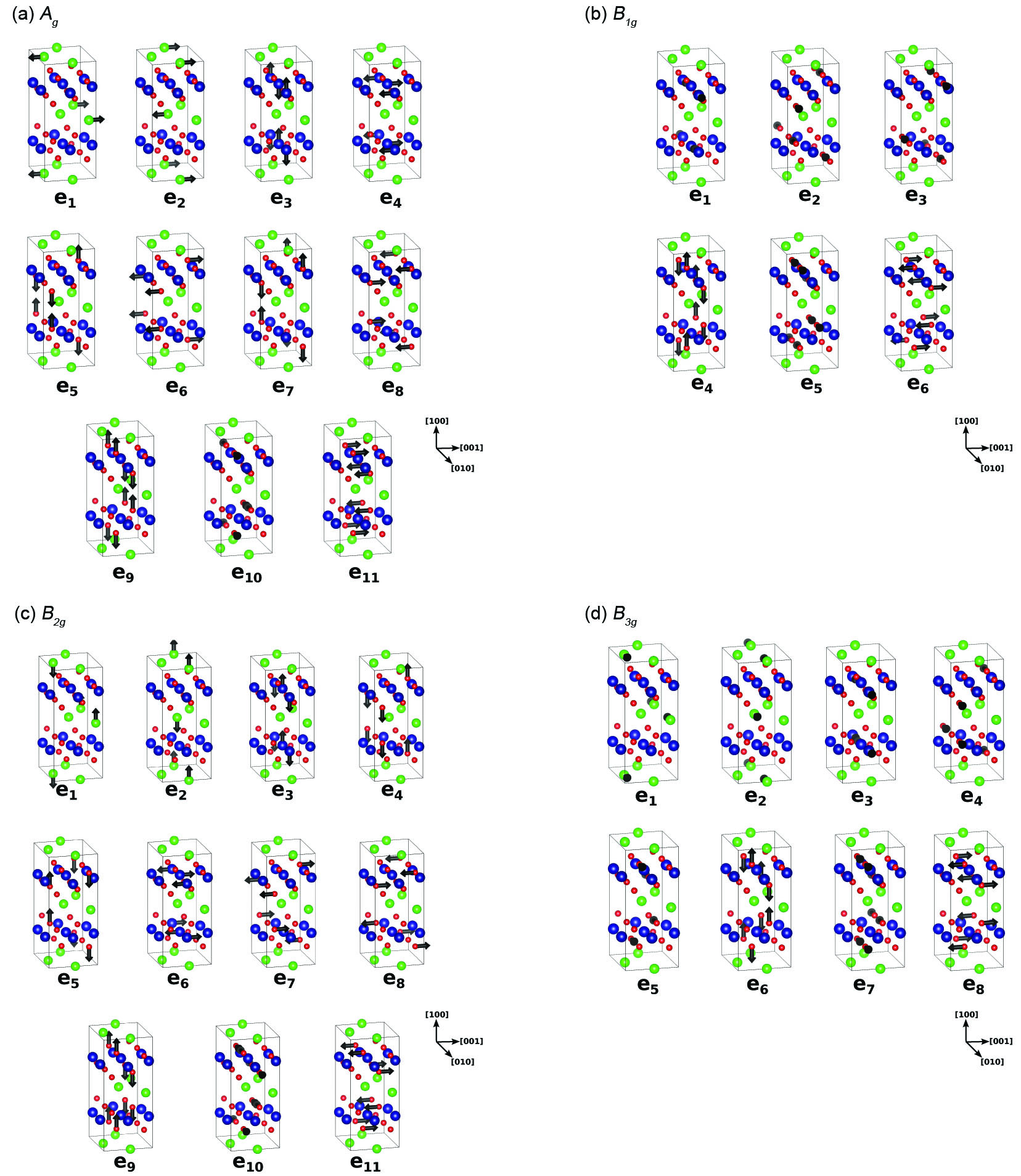}
 \end{center}
 \caption{Symmetry adapted modes of the irreps $A_g$, $B_{1g}$, $B_{2g}$, and $B_{3g}$. Blue, red, and green atoms denote Cr$^{3+}$, O$^{2-}$, and Sr$^{2+}$ ions, respectively. The ionic radii in the figure are changed for clarity.}
    \label{fig:SymMod}
\end{figure*}

In \sco, the four Raman active irreducible representations (irreps) at the $\Gamma$-point are $A_g$, $B_{1g}$, $B_{2g}$, and $B_{3g}$ with Raman tensors given in (\ref{eqn:ramtens}). In this section, we present the frequencies and eigendisplacements for each of the Raman active phonon modes. Tables~\ref{table:SymModAg}, \ref{table:SymModB1g}, \ref{table:SymModB2g}, and \ref{table:SymModB3g} list the phonon frequencies calculated from first principles and the phonon eigendisplacements expressed in the basis of the ionic displacements shown in the corresponding Figures~\ref{fig:SymMod}(a)-(d).  The results presented in these tables are obtained from calculations assuming ferromagnetic ordering and a Hubbard $U$ of 3~eV, the effects of which are discussed further in Sec.~\ref{sec:magconf} and \ref{sec:HubbU}, respectively.

\begin{table*}
\centering{}
\begin{tabular}{|c|c|}
\hline
Frequency & Eigendisplacement \\
(\cm) & \\ \hline
100 & (0.066, -0.082, 0.005, -0.009, -0.017, -0.012, 0.019, 0.003, -0.001, -0.011, -0.014)\\
142 & (-0.080, -0.070, -0.005, 0.008, 0.012, 0.018, 0.003, 0.018, -0.005, 0.005, 0.002)\\
249 & (-0.011, 0.002, 0.119, -0.012, -0.053, -0.032, 0.071, -0.051, -0.007, -0.063, 0.011)\\
310 & (0.010, 0.001, 0.029, 0.122, 0.001, 0.070, -0.028, 0.071, 0.000, -0.021, 0.009)\\
413 & (-0.003, -0.001, -0.022, -0.011, -0.138, 0.031, -0.046, 0.020, 0.150, -0.085, 0.092)\\
428 & (-0.002, 0.004, -0.057, 0.021, -0.021, 0.035, 0.154, -0.070, -0.074, -0.121, 0.016)\\
454 & (-0.007, 0.004, -0.015, 0.012, -0.003, -0.109, -0.001, 0.062, 0.042, -0.119, -0.172)\\
469 & (0.002, -0.008, -0.001, 0.030, 0.116, -0.106, -0.076, -0.117, 0.046, -0.048, 0.103)\\
528 & (0.004, 0.001, 0.014, -0.047, 0.087, 0.060, -0.070, 0.102, -0.081, -0.138, 0.051)\\
581 & (-0.004, -0.002, -0.012, 0.017, -0.137, -0.098, -0.098, -0.011, -0.148, 0.004, 0.029)\\
600 & (0.000, -0.004, 0.005, -0.006, -0.019, 0.118, -0.117, -0.144, -0.015, -0.041, -0.108)\\
\hline
\end{tabular}
\caption{The phonon frequencies and the corresponding eigendisplacements for the Raman active irrep $A_g$. The dynamical matrix eigenvectors are normalized, and the eigendisplacements are given in the basis of the displacements shown in Fig. \ref{fig:SymMod}(a).}
\label{table:SymModAg}
\end{table*}

\begin{table*}
\centering{}
\begin{tabular}{|c|c|}
\hline
Frequency & Eigendisplacement \\
(\cm) & \\ \hline
372&(0.113, 0.037, 0.050, -0.109, -0.039, -0.059)\\
424&(-0.024, 0.040, -0.010, -0.152, 0.188, 0.023)\\
440&(-0.026, -0.189, 0.130, -0.058, -0.002, -0.064)\\
462&(0.040, -0.020, 0.119, 0.060, 0.052, 0.192)\\
473&(-0.046, 0.039, 0.016, -0.138, -0.152, 0.110)\\
569&(-0.038, 0.148, 0.169, 0.039, 0.003, -0.077)\\
\hline
\end{tabular}
\caption{The phonon frequencies and the corresponding eigendisplacements for the Raman active irrep $B_{1g}$. The dynamical matrix eigenvectors are normalized, and the eigendisplacements are given in the basis of the displacements shown in Fig. \ref{fig:SymMod}(b).}
\label{table:SymModB1g}
\end{table*}

\begin{table*}
\centering{}
\begin{tabular}{|c|c|}
\hline
Frequency & Eigendisplacement \\
(\cm) & \\ \hline
146&(0.073, 0.067, -0.046, -0.029, -0.031, -0.007, -0.015, -0.009, -0.002, 0.013, -0.003)\\
221&(0.075, -0.069, 0.027, -0.010, 0.031, -0.005, -0.013, -0.009, -0.033, -0.016, 0.003)\\
263&(-0.015, -0.042, -0.104, -0.046, -0.061, -0.020, -0.039, -0.065, -0.018, 0.057, -0.009)\\
308&(-0.006, 0.010, 0.038, 0.006, -0.028, -0.120, -0.069, -0.066, 0.001, -0.027, 0.012)\\
409&(-0.001, 0.005, -0.014, 0.147, -0.047, 0.021, -0.041, -0.030, -0.147, -0.064, 0.092)\\
426&(-0.001, -0.003, -0.057, 0.039, 0.152, -0.025, -0.052, 0.082, 0.064, -0.100, 0.043)\\
453&(-0.003, -0.001, -0.018, -0.032, -0.026, -0.012, 0.107, -0.021, -0.054, -0.169, -0.125)\\
463&(0.004, 0.005, 0.000, 0.118, 0.078, 0.019, -0.057, -0.116, 0.031, 0.039, -0.148)\\
515&(0.002, 0.001, -0.014, 0.067, 0.056, -0.052, 0.111, 0.092, -0.097, 0.117, -0.042)\\
583&(0.011, -0.007, -0.013, 0.086, -0.022, -0.012, 0.151, -0.088, 0.121, -0.003, 0.088)\\
592&(-0.009, 0.014, 0.003, -0.098, 0.143, 0.007, 0.044, -0.128, -0.085, 0.013, 0.071)\\
\hline
\end{tabular}
\caption{The phonon frequencies and the corresponding eigendisplacements for the Raman active irrep $B_{2g}$. The dynamical matrix eigenvectors are normalized, and the eigendisplacements are given in the basis of the displacements shown in Fig. \ref{fig:SymMod}(c).}
\label{table:SymModB2g}
\end{table*}

\begin{table*}
\centering{}
\begin{tabular}{|c|c|}
\hline
Frequency & Eigendisplacement \\
(\cm) & \\ \hline
83&(0.068, -0.082, -0.003, 0.001, -0.002, 0.000, 0.019, -0.005)\\
149&(-0.082, -0.068, -0.009, 0.006, 0.003, 0.007, -0.003, 0.001)\\
373&(-0.003, -0.006, 0.113, 0.038, 0.052, -0.109, -0.036, -0.058)\\
426&(-0.005, 0.004, -0.019, 0.064, -0.030, -0.133, 0.195, 0.024)\\
442&(0.001, 0.000, 0.028, 0.181, -0.138, 0.075, -0.026, 0.043)\\
462&(-0.002, 0.002, -0.024, -0.010, -0.104, -0.014, -0.002, -0.223)\\
476&(-0.004, 0.002, 0.057, -0.046, 0.023, 0.159, 0.149, -0.040)\\
571&(0.003, 0.004, -0.038, 0.148, 0.169, 0.041, 0.003, -0.075)\\
\hline
\end{tabular}
\caption{The phonon frequencies and the corresponding eigendisplacements for the Raman active irrep $B_{3g}$. The dynamical matrix eigenvectors are normalized, and the eigendisplacements are given in the basis of the displacements shown in Fig. \ref{fig:SymMod}(d).}
\label{table:SymModB3g}
\end{table*}

\subsection{Dependence of $\Gamma$-point phonons on the magnetic configuration}
\label{sec:magconf}

\begin{figure}
 \begin{center}
 \includegraphics[width=8cm]{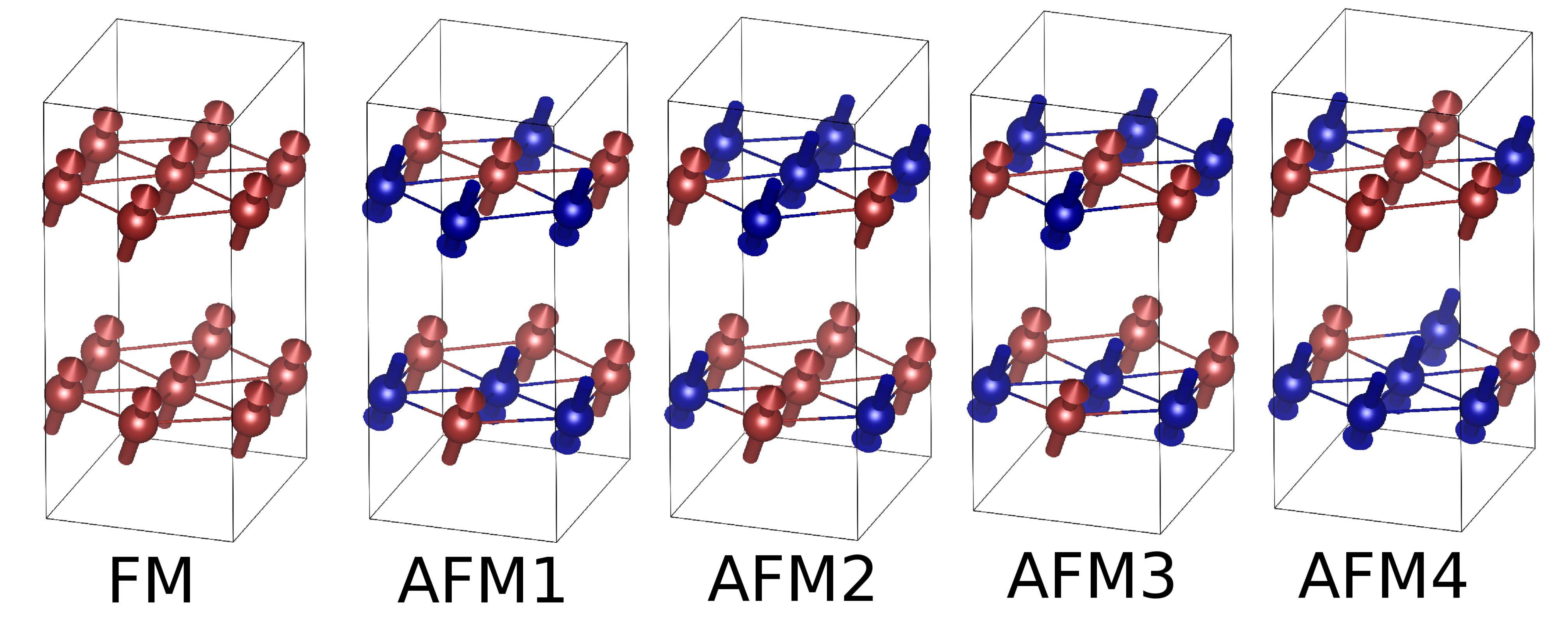}
 \end{center}
 \caption{The five different magnetic configurations used to calculate the phonon frequencies. For simplicity, only the Cr$^{3+}$ ions are included in the figure, and opposite spin Cr$^{3+}$ ions are represented with different colors.}
    \label{fig:MagneticOrders}
\end{figure}

In order to elucidate the effect of the magnetic order on the phonon frequencies, we calculated the $\Gamma$-point phonon frequencies considering ferromagnetic (FM) and the various antiferromagnetic (AFM) configurations shown in Fig.~\ref{fig:MagneticOrders}. The crystal structure determined from neutron powder diffraction\cite{Dutton2011} was relaxed in the FM state with the lattice vectors fixed and was then used for the phonon calculations in each of the different magnetic configurations.  For calculations in the AFM configurations, the zero point forces were subtracted. The results for $U\!=\!3$~eV are presented in Table~\ref{table:AllGammaFreqs}. All of the optical phonon modes that showed a frequency shift of $\geq$15~\cm\ are in the range of 267 to 402~\cm\ (indicated by $\ast$) which coincides with the range over which the higher-frequency magnetic excitation centered at 310~\cm\ (38~meV) was observed. Of the twelve modes showing significant spin-phonon coupling, four are Raman active as discussed in Sec.~\ref{sec:RamActPhonons}.

\begin{table*}
\centering{}
\begin{tabular}{|cc|ccccc||cc|ccccc|}
\hline
Irrep	&		&	FM	&	AFM1	&	AFM2	&	AFM3	&	AFM4	 &Irrep	&		&	FM	&	AFM1	&	AFM2	&	AFM3	&	AFM4	\\
\hline
$A_u$ $(\Gamma_1^-)$	&	Silent	        &	47	&	48	&	47	&	47	&	47	&	$A_g$ $(\Gamma_1^+)$	&	Raman active	&	428	&	427	&	428	&	427	&	428	\\
$B_{3u}$ $(\Gamma_3^-)$	&	IR active (a)	&	73	&	73	&	73	&	72	&	73	&	$A_u$ $(\Gamma_1^-)$	&	Silent	        &	428	&	430	&	425	&	430	&	425	\\
$B_{3g}$ $(\Gamma_3^+)$	&	Raman active	&	83	&	83	&	83	&	83	&	83	&	$B_{1g}$ $(\Gamma_2^+)$	&	Raman active	&	440	&	441	&	440	&	441	&	440	\\
$A_g$ $(\Gamma_1^+)$	&	Raman active	&	100	&	100	&	100	&	100	&	100	&	$B_{3g}$ $(\Gamma_3^+)$	&	Raman active	&	442	&	443	&	442	&	443	&	442	\\
$B_{2u}$ $(\Gamma_4^-)$	&	IR active (b)	&	101	&	101	&	101	&	101	&	101	&	$B_{3u}$ $(\Gamma_3^-)$	&	IR active (a)	&	451	&	449	&	450	&	449	&	450	\\
$B_{1u}$ $(\Gamma_2^-)$	&	IR active (c)	&	126	&	125	&	124	&	125	&	124	&	$B_{2g}$ $(\Gamma_4^+)$	&	Raman active	&	453	&	448	&	450	&	448	&	450	\\
$B_{1u}$ $(\Gamma_2^-)$	&	IR active (c)	&	134	&	132	&	132	&	132	&	132	&	$A_g$ $(\Gamma_1^+)$	&	Raman active	&	454	&	451	&	453	&	451	&	453	\\
$B_{2u}$ $(\Gamma_4^-)$	&	IR active (b)	&	138	&	138	&	138	&	138	&	138	&	$A_u$ $(\Gamma_1^-)$	&	Silent	        &	454	&	459	&	454	&	459	&	454	\\
$A_g$ $(\Gamma_1^+)$	&	Raman active	&	142	&	142	&	142	&	142	&	142	&	$B_{2u}$ $(\Gamma_4^-)$	&	IR active (b)	&	455	&	460	&	455	&	460	&	455	\\
$B_{2g}$ $(\Gamma_4^+)$	&	Raman active	&	146	&	146	&	146	&	146	&	146	&	$B_{1u}$ $(\Gamma_2^-)$	&	IR active (c)	&	457	&	456	&	457	&	456	&	457	\\
$B_{3g}$ $(\Gamma_3^+)$	&	Raman active	&	149	&	148	&	149	&	148	&	149	&	$B_{3g}$ $(\Gamma_3^+)$	&	Raman active	&	462	&	462	&	462	&	462	&	462	\\
$B_{3u}$ $(\Gamma_3^-)$	&	IR active (a)	&	159	&	155	&	155	&	155	&	155	&	$B_{1g}$ $(\Gamma_2^+)$	&	Raman active	&	462	&	462	&	461	&	462	&	461	\\
$B_{1u}$ $(\Gamma_2^-)$	&	IR active (c)	&	174	&	174	&	174	&	174	&	174	&	$B_{2g}$ $(\Gamma_4^+)$	&	Raman active	&	463	&	460	&	460	&	460	&	460	\\
$B_{3u}$ $(\Gamma_3^-)$	&	IR active (a)	&	221	&	221	&	221	&	221	&	221	&	$B_{3u}$ $(\Gamma_3^-)$	&	IR active (a)	&	463	&	463	&	462	&	463	&	462	\\
$B_{2g}$ $(\Gamma_4^+)$	&	Raman active	&	221	&	221	&	221	&	221	&	221	&	$A_u$ $(\Gamma_1^-)$	&	Silent	        &	465	&	472	&	470	&	472	&	470	\\
$A_u$ $(\Gamma_1^-)$	&	Silent	        &	226	&	223	&	223	&	223	&	223	&	$A_g$ $(\Gamma_1^+)$	&	Raman active	&	469	&	464	&	465	&	464	&	465	\\
$B_{2u}$ $(\Gamma_4^-)$	&	IR active (b)	&	239	&	233	&	233	&	233	&	233	&	$B_{2u}$ $(\Gamma_4^-)$	&	IR active (b)	&	469	&	476	&	473	&	476	&	473	\\
$A_g$ $(\Gamma_1^+)$	&	Raman active	&	249	&	249	&	248	&	249	&	248	&	$B_{1g}$ $(\Gamma_2^+)$	&	Raman active	&	473	&	473	&	475	&	473	&	475	\\
$B_{3u}$ $(\Gamma_3^-)$	&	IR active (a)	&	257	&	254	&	255	&	254	&	254	&	$B_{1u}$ $(\Gamma_2^-)$	&	IR active (c)	&	476	&	472	&	472	&	472	&	472	\\
$B_{1u}$ $(\Gamma_2^-)$	&	IR active (c)	&	260	&	256	&	256	&	256	&	256	&	$B_{3g}$ $(\Gamma_3^+)$	&	Raman active	&	476	&	478	&	478	&	478	&	478	\\
$B_{2g}$ $(\Gamma_4^+)$	&	Raman active	&	263	&	264	&	262	&	264	&	262	&	$B_{3u}$ $(\Gamma_3^-)$	&	IR active (a)	&	480	&	483	&	476	&	483	&	476	\\
$\ast A_u$ $(\Gamma_1^-)$	&	Silent	        &	267	&	248	&	246	&	248	&	246	&	$B_{1u}$ $(\Gamma_2^-)$	&	IR active (c)	&	490	&	491	&	487	&	491	&	487	\\
$\ast B_{2u}$ $(\Gamma_4^-)$	&	IR active (b)	&	268	&	251	&	249	&	251	&	250	&	$B_{2u}$ $(\Gamma_4^-)$	&	IR active (b)	&	492	&	490	&	489	&	490	&	489	\\
$\ast B_{2u}$ $(\Gamma_4^-)$	&	IR active (b)	&	306	&	287	&	287	&	287	&	287	&	$A_u$ $(\Gamma_1^-)$	&	Silent	        &	493	&	491	&	491	&	491	&	491	\\
$\ast A_u$ $(\Gamma_1^-)$	&	Silent	        &	306	&	287	&	287	&	287	&	287	&	$B_{2g}$ $(\Gamma_4^+)$	&	Raman active	&	515	&	513	&	515	&	513	&	515	\\
$\ast B_{2g}$ $(\Gamma_4^+)$	&	Raman active	&	308	&	292	&	292	&	292	&	292	&	$B_{3u}$ $(\Gamma_3^-)$	&	IR active (a)	&	517	&	523	&	517	&	523	&	517	\\
$\ast A_g$ $(\Gamma_1^+)$	&	Raman active	&	310	&	293	&	293	&	293	&	293	&	$B_{1u}$ $(\Gamma_2^-)$	&	IR active (c)	&	519	&	526	&	520	&	526	&	520	\\
$\ast B_{1g}$ $(\Gamma_2^+)$	&	Raman active	&	372	&	303	&	367	&	303	&	367	&	$B_{3u}$ $(\Gamma_3^-)$	&	IR active (a)	&	526	&	527	&	527	&	527	&	527	\\
$\ast B_{3g}$ $(\Gamma_3^+)$	&	Raman active	&	373	&	305	&	368	&	305	&	368	&	$A_g$ $(\Gamma_1^+)$	&	Raman active	&	528	&	527	&	528	&	527	&	528	\\
$\ast B_{3u}$ $(\Gamma_3^-)$	&	IR active (a)	&	382	&	321	&	321	&	321	&	321	&	$B_{1u}$ $(\Gamma_2^-)$	&	IR active (c)	&	545	&	545	&	545	&	545	&	545	\\
$\ast B_{1u}$ $(\Gamma_2^-)$	&	IR active (c)	&	384	&	323	&	322	&	323	&	322	&	$B_{3u}$ $(\Gamma_3^-)$	&	IR active (a)	&	551	&	553	&	554	&	553	&	554	\\
$\ast B_{3u}$ $(\Gamma_3^-)$	&	IR active (a)	&	400	&	370	&	367	&	370	&	367	&	$B_{1g}$ $(\Gamma_2^+)$	&	Raman active	&	569	&	559	&	561	&	559	&	561	\\
$\ast B_{1u}$ $(\Gamma_2^-)$	&	IR active (c)	&	402	&	371	&	368	&	371	&	368	&	$B_{3g}$ $(\Gamma_3^+)$	&	Raman active	&	571	&	561	&	563	&	561	&	563	\\
$A_u$ $(\Gamma_1^-)$	&	Silent	        &	404	&	406	&	406	&	406	&	406	&	$A_g$ $(\Gamma_1^+)$	&	Raman active	&	580	&	576	&	576	&	576	&	576	\\
$B_{2g}$ $(\Gamma_4^+)$	&	Raman active	&	409	&	408	&	409	&	408	&	409	&	$B_{1u}$ $(\Gamma_2^-)$	&	IR active (c)	&	581	&	573	&	577	&	573	&	577	\\
$B_{2u}$ $(\Gamma_4^-)$	&	IR active (b)	&	410	&	409	&	409	&	409	&	409	&	$B_{2g}$ $(\Gamma_4^+)$	&	Raman active	&	583	&	579	&	580	&	579	&	580	\\
$A_g$ $(\Gamma_1^+)$	&	Raman active	&	413	&	413	&	413	&	412	&	413	&	$B_{3u}$ $(\Gamma_3^-)$	&	IR active (a)	&	591	&	586	&	587	&	586	&	587	\\
$B_{1g}$ $(\Gamma_2^+)$	&	Raman active	&	424	&	428	&	427	&	428	&	427	&	$B_{2g}$ $(\Gamma_4^+)$	&	Raman active	&	592	&	587	&	587	&	587	&	587	\\
$B_{2g}$ $(\Gamma_4^+)$	&	Raman active	&	426	&	425	&	427	&	425	&	427	&	$A_g$ $(\Gamma_1^+)$	&	Raman active	&	600	&	593	&	594	&	593	&	594	\\
$B_{3g}$ $(\Gamma_3^+)$	&	Raman active	&	426	&	429	&	428	&	429	&	428	&	$B_{1u}$ $(\Gamma_2^-)$	&	IR active (c)	&	642	&	643	&	642	&	643	&	642	\\
$B_{2u}$ $(\Gamma_4^-)$	&	IR active (b)	&	427	&	432	&	426	&	432	&	426	&&&&&&&	\\													
\hline
\end{tabular}
\caption{The 81 optical phonon frequencies on the $\Gamma$-point, calculated from first principles with the frozen phonons method in \cm. The $U$ value used for the DFT+$U$ calculations is 3~eV. The magnetic orders refer to the states presented in Fig.~\ref{fig:MagneticOrders}. Phonons showing frequency shifts of $\geq$15~\cm\ for the different magnetic configurations are indicated with an asterisk ($\ast$).}
\label{table:AllGammaFreqs}
\end{table*}

\subsection{Dependence of the spin-phonon coupling on the Hubbard $U$ in DFT+$U$}
\label{sec:HubbU}

The value of the Hubbard $U$ used in the DFT+$U$ calculations presented in the main text was chosen to be 3~eV. This value has been previously shown to correctly reproduce the physics of different Cr compounds.\cite{Fennie2005,Fennie2006} In order to show that the physical results obtained from our calculations are robust against variations of this parameter, we repeated the phonon calculations with different values of $U$, namely for $U\!=\!2$~eV and $U\!=\!4$~eV. In Table~\ref{table:UAg}, the frequencies of the Raman active $A_g$ $\Gamma$-point phonons are listed for these values of $U$ and the five different magnetic configurations considered. Changing the value of $U$ causes changes in the hybridization of the Cr$^3+$ ions with their neighbors and so has quantitative effects on the phonon frequencies. In particular, the frequencies change as much as few percent when $U$ is changed by 1~eV. However, the primary qualitative result that is mentioned in the main text, i.e.\@ that only one $A_g$ mode has strong spin-phonon coupling (defined by a change of phonon frequency by $\geq\!15$~\cm\ in different magnetic states), is independent of the value of $U$ chosen. While we do not present the results for other phonon modes, our calculations indicate that the same result applies to all $\Gamma$-point phonons, and hence the qualitative conclusions are not dependent on the value of $U$ chosen.

\begin{table*}[t!]
%\centering{}
\begin{tabular}{|ccccc|ccccc|ccccc|}
\hline
\multicolumn{5}{|c|}{$U\!=\!2$~eV} & \multicolumn{5}{c}{$U\!=\!3$~eV} & \multicolumn{5}{|c|}{$U\!=\!4$~ eV} \\
FM& AFM1& AFM2& AFM3& AFM4& FM& AFM1& AFM2& AFM3& AFM4& FM& AFM1& AFM2& AFM3& AFM4\\
\hline
101& 100& 100& 100& 100& 100& 100& 100& 100& 100& 101& 101& 101& 101& 101\\
142& 141& 141& 141& 141& 142& 142& 142& 142& 142& 142& 142& 142& 142& 142\\
247& 247& 246& 248& 246& 249& 249& 248& 249& 248& 251& 251& 250& 251& 250\\
308& 286& 286& 286& 286& 310& 293& 293& 293& 293& 313& 299& 299& 299& 299\\
406& 407& 408& 407& 408& 413& 413& 413& 412& 413& 417& 417& 418& 417& 418\\
420& 417& 419& 417& 419& 428& 427& 428& 427& 428& 435& 435& 436& 435& 436\\
449& 444& 446& 444& 446& 454& 451& 453& 451& 453& 460& 457& 459& 458& 459\\
464& 458& 459& 458& 459& 469& 464& 465& 464& 465& 474& 470& 471& 470& 471\\
519& 517& 519& 517& 519& 528& 527& 528& 527& 528& 537& 536& 537& 536& 537\\
573& 570& 570& 570& 570& 580& 576& 576& 576& 576& 584& 582& 581& 582& 583\\
592& 584& 585& 584& 585& 600& 593& 594& 593& 594& 605& 601& 601& 601& 601\\
\hline
\end{tabular}
\caption{$A_g$ phonon frequencies from first principles, calculated for different magnetic configurations and different values of Hubbard $U$ used in the DFT+$U$ calculations, in \cm. While changing the value of $U$ causes minor quantitative changes, the fact that only the mode at $\sim\!300$~\cm\ has strong spin-phonon coupling is robust against variations of $U$.}
\label{table:UAg}
\end{table*}

\clearpage

\bibliography{./SrCr2O4}

\end{document}